\documentclass{jkas}

\def\year{2015} 
\def\month{October} 
\def\volume{00} 
\def\issue{0} 
\def\beginpage{1} 
\def\endpage{99} 
\def\received{July xx, 2015} 
\def\accepted{September xx, 2015} 
\date{Received \received; accepted \accepted}

\input colordvi

\title{
Amplitude Correction Factors Of Korean VLBI Network Observations
\thanks{Part of a special issue on the Korean VLBI Network (KVN)}
}

\author[1,2]{Sang-Sung Lee\thanks{sslee@kasi.re.kr}}
\author[1]{Do-Young Byun}
\author[1]{Chung Sik Oh}
\author[1]{Hyo Ryoung Kim}
\author[1,2]{Jongsoo Kim}
\author[1,2]{Taehyun Jung}
\author[1]{Se-Jin Oh}
\author[1]{Duk-Gyoo Roh}
\author[1]{Dong-Kyu Jung}
\author[1]{Jae-Hwan Yeom}

\affil[1]{Korea Astronomy and Space Science Institute,
776 Daedeok-daero, Yuseong-gu, Daejeon 34055,
Republic of Korea;\email{sslee@kasi.re.kr}}
\affil[2]{Korea University of Science and Technology,
217 Gajeong-ro, Yuseong-gu, Daejeon 34113,
Republic of Korea}

\def\corrauthor{
S.-S. Lee
}

\def\runningauthor{
Lee et al.
}

\def\runningtitle{
Amplitude Correction Factors of KVN Observations
}

\def\keywords{
Techniques: interferometric --- Instrumentation: interferometers
--- Radio continuum: galaxies
}

\def\abstracttext{
We report results of investigation of amplitude calibration
for very long baseline interferometry (VLBI) observations
with Korean VLBI Network (KVN).
Amplitude correction factors are estimated based on 
comparison of KVN observations at 22~GHz correlated
by Daejeon hardware correlator and DiFX software correlator
in Korea Astronomy and Space Science Institute (KASI)
with Very Long Baseline Array (VLBA)
observations at 22~GHz by DiFX software correlator in
National Radio Astronomy Observatory (NRAO).
We used the observations for compact radio sources,
3C~454.3, NRAO~512, OJ 287, BL Lac, 3C 279, 1633+382, and 1510$-$089,
which are almost unresolved for baselines
in a range of 350-477~km.
Visibility data of the sources obtained with similar baselines
at KVN and VLBA are selected, fringe-fitted, calibrated, and
compared for their amplitudes.
We found that visibility amplitudes of KVN observations
should be corrected by factors of 1.10 and 1.35
when correlated by DiFX and Daejeon correlators, respectively.
These correction factors are attributed to
the combination of two steps of 2-bit quantization in KVN observing systems
and characteristics of Daejeon correlator.
}

\begin{document}
\jkashead 


\section{Introduction\label{sec:intro}}

The Korean VLBI Network (KVN) was built in Korea,
as the first mm-dedicated Very Long Baseline Interferometry (VLBI) network,
with its main goals for high resolution multi-frequency study
of the formation and death of stars,
the structure and dynamics of our own Galaxy,
and the nature of active galactic nuclei (AGNs)~\citep{lee+11,lee+14}. 
The KVN consists of three 21-m radio telescopes:
KVN Yonsei Radio Telescope (KY),
KVN Ulsan Radio Telescope (KU),
and KVN Tanma Radio Telescope (KT),
with baseline lengths of 305-477~km,
operating at four frequency bands of 22~GHz, 43~GHz, 86~GHz, and 129~GHz.
The maximum angular resolution at the 129~GHz band is 
$\sim$ 1\,milliarcsecond (mas).

Amplitude calibration of the KVN observations is performed
according to a conventional VLBI calibration
consisting of (a) calibration for amplitude variation
due to atmospheric fluctuation, receiver noise fluctuation,
elevation dependence of the antenna gains, etc,
(b) amplitude correction for the atmospheric opacity on an antenna,
and (c) re-normalization of the fringe amplitude for restoring
amplitude distortion due to quantization~\citep{lee+14}.
In order to confirm the amplitude calibration of
the KVN observations,
the VLBI data correlated with DiFX correlator
were compared with that of the Very Long Baseline Array (VLBA)
observations conducted on the same source
and in a similar time, assuming the amplitude calibration of the VLBA
has been confirmed.
\cite{lee+14} compared the KVN 22~GHz observations conducted on 2010 October 1
with the VLBA observations on 2010 October 2. 
They selected the {\it uv}-data with the same {\it uv}-distances from both
observations for extragalactic compact radio sources,
and compared them each other, estimating the correction factors
of $<$ 13\% for individual antenna.
\cite{pet+12} found that the amplitude correction factors of KVN observations
at 43~GHz are within 30\% based on the simlar comparison with
VLBA observations.
However, the authors for the previous works could not conclude
to which these amplitude correction factors were specifically attributed
and whether they should be applied to all KVN observations or not.

In this paper we report
our understanding of the KVN systems related to the amplitude calibration,
 our findings for the KVN amplitude correction factors for the data
correlated by DiFX and Daejeon correlators, respectively,
and the reasonable suggestion of the application of the correction factors
to the future KVN observations, based on observational tests 
of re-quantization losses in the KVN system.
In Section 2, 
we overview general aspects of the KVN observing systems
focusing on the two-step quantization loss.
In Section 3, we summarize the observation data used and
describe the data reduction procedure.
In Section 4, we report the results of the amplitude correction factors.
Finally, we make discussion and conclusion in Section 5.

\section{Overview of the KVN observing systems}

\begin{figure*}[!t]
        \centering 
        \includegraphics[width=160mm]{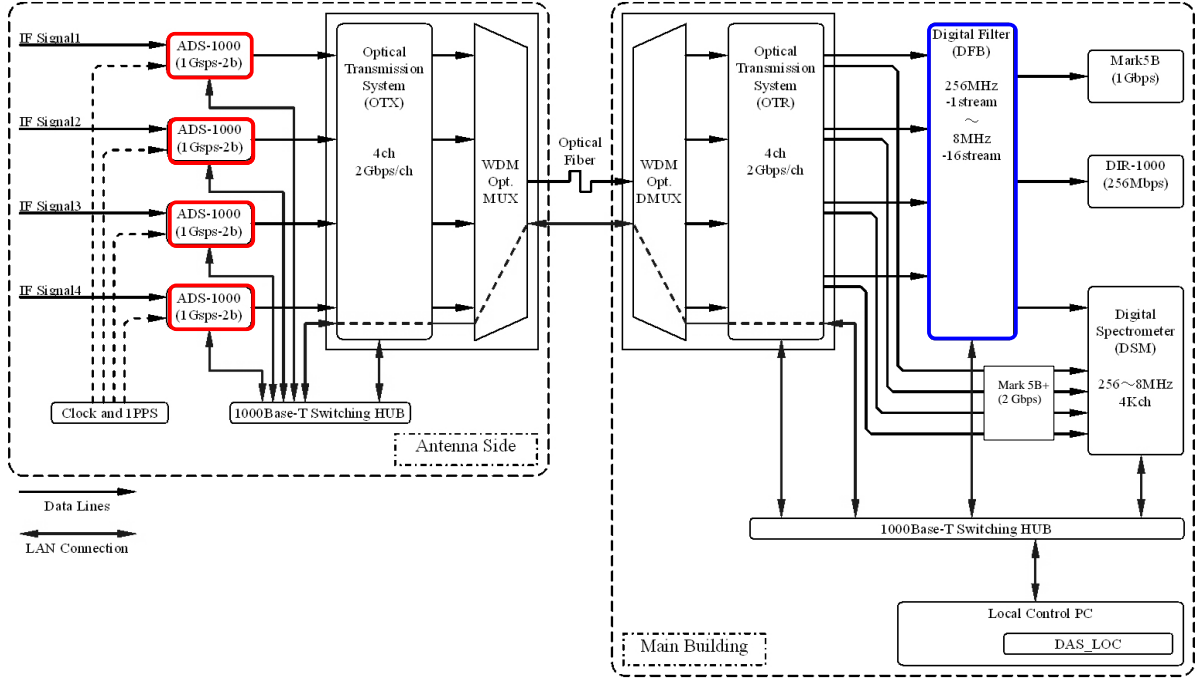}
        \caption{ 
        Schematic diagram of KVN Data Acquisition System. Adopted and modified from \cite{oh+11}.
	The digital samplers and the DFB are outlined in red and blue, respectively.
        }
        \label{fig1} 
\end{figure*}

The KVN observing system consists of
three 21-m antennas,
quasi-optical systems,
receivers operating at four frequency bands
-- 22~GHz, 43~GHz, 86~GHz, and 129~GHz --
and backend systems, as described in \cite{lee+11}.
The backend systems include the KVN Data Acquisition System (DAS)
which consists of four subsystems:
digital samplers, optical transmission system (OTS),
digital filter bank (DFB) and digital spectrometer (DSM),
as shown in Figure~\ref{fig1}.
The digital samplers in antenna cabin digitize the observation data
into 2-bit data streams with four quantization levels.
The OTS transmits the digitized signals to the observing building
through optical fibers.
The digitized signals are fed into the digital filter bank (DFB),
which process the signals to be
16 sub-bands of 16~MHz wide with a frequency interval of 16~MHz.
The processed signals are recorded by the Mark 5B system
at an aggregate recording rate of 1024~Mbps.
For the observations at the aggregate recording rate of 2048~Mbps,
the digitized signals from the digital samplers
can be directly recorded into the Mark 5B+ system
with a single band of 512~MHz bandwidth,
without passing through the DFB.
The recorded data are shipped to
the Korea-Japan Correlation Center (KJCC) and correlated
with the DiFX software correlator or the Daejeon hardware correlator.
In this section, we focus on describing the digital sampler
and the digital filter bank systems, which generate the quantization loss
of the data, 
although detailed information for the KVN DAS is described in \cite{oh+11},
and on introducing the DiFX software correlator and 
Daejeon hardware correlator.

\subsection{Digital sampler and Digital filter bank}\label{digital}

In VLBI systems, the digital sampler is more preferable to the analog
sampler due to important practical advantages for compensating
time delays and measuring cross-correlation of signals such as~\citep{tho+01}:
(a) higher accuracy ($\sim$10-100 psec) of the long delay can be easily
obtained in digital signal processing, and
(b) the digital signal is not distorted by the digital instrumentation
except for the quantization whose effects can be estimated.
The digital sampler performs sampling the voltages of the analog signals
at periodic intervals and quantizing the sampled value of the voltages
with a small number of bits (e.g., 1-bit or 2-bit). The low level
(or small bit number) quantization result in a quantization noise, i.e.,
a loss of sensitivity in digitizing the analog signals.
Therefore, the optimal number of quantization levels is important
in the digital signal process,
and the investigations of the optimal quantization
levels can be found
in the literature~\citep[][and references therein]{tho+01}.
The investigations result in the well-known efficiency factors,
for Nyquist sampling, 0.637 for 2-level quantization,
0.810 for 3-level, 
and 0.881 for 4-level.
It should be noted that the efficiency factors
for 3-level and 4-level quantizations are
the maximum values when the sampling threshold is properly set,
and hence, in practice, lower efficiencies are expected if the sampling
threshold is not properly adjusted.

The digital samplers used in KVN is based on
ADS-1000 Giga-Bit-Sampler (GBS) developed by the 
National Institute of Information
and Communications Technology (NiCT), Japan~\citep{nak+00}
The digital sampler quantizes an intermediate frequency analog signal,
converted from the radio frequency signals by receivers,
with a bandwidth of 512~MHz
and in four quantization levels (or in two bits),
in order to achieve a higher sensitivity in the quantization.
Since the bandwidth of the sampled signals by the digital sampler is 512~MHz,
it is necessary to process the signals fitted into a bandwidth
of 256~MHz for the Mark 5B system.
Moreover, it is also required to channelize the whole frequency band
into sub-bands and pick up the desired sub-bands optimized for
scientific interests.
Therefore, the DFB is introduced in the KVN DAS
for flexibly channelizing and extracting the desired
frequency bands according to the scientific purposes.
In fact, the DFB filters out the desired frequency bands
in the digital waveforms generated
by the digital samplers with four-level (2-bit) quantized sampling.
Since the data processing rate is limited
by the instrumental capability,
it is required to re-quantize the signal
after the digital filtering~\citep{igu+05}.

\subsection{DiFX and Daejeon correlator}\label{correlator}

A software correlator based on DiFX~\citep[Distributed FX;][]{del+07,del+11}
has been installed and used mainly
for correlating data from the KVN observations.
The software version of DiFX has been
upgraded and the current version is 2.3.
Detailed information of the DiFX installed for the KVN is described
in \cite{lee+14}.
A new hardware correlator (Daejeon correlator) was developed
in 2009 by Korea Astronomy and Space Science Institute (KASI) 
and National Astronomical Observatory of Japan (NAOJ).
The Daejeon correlator is mainly used 
for the East Asian VLBI Network (EAVN) consisting of
Korean VLBI Network (KVN),
Japanese VLBI Network (JVN) including  
the VLBI Exploration of Radio Astrometry (VERA) in Japan,
and the Chinese VLBI Network (CVN).
More information of the Daejeon correlator is 
summarized in \cite{lee+14}.

In order to evaluate the performance of the new hardware correlator,
\cite{lee+15} carefully compared the correlation outputs of
both the DiFX and Daejeon correlators,
using the KVN observations conducted at 22~GHz on 2011 January 28-29.
They found that the two correlators are comparable each other
in the correlation outputs except for the visibility amplitudes
of the hardware correlator being lower by a factor of $<$8\%
than those of the DiFX correlator.
The amplitude difference is due to the characteristics of
the hardware correlator: (a) the way of fringe phase tracking (causing
an amplitude loss of $\sim5\%$ due to coarse phase tracking and see \cite{igu+00})
and (b) an unusual pattern of the amplitude of the correlation output
(causing an amplitude loss of $\sim3\%$).
Fortunately, it was announced by the KJCC operation team
that the latter characteristic has been fixed for the data correlated
after 2015 March.

It should be noted that \cite{lee+15} did not
apply digital correction (by switching off the option DIGICOR of the task FITLD)
in order to compare the raw outputs from the two correlators.
In fact, the option is required for proper amplitude calibration
in case of DiFX data, usually, 
unless DiFX was configured to use Tsys files during correlation.
However, for the case of the hardware correlator, the option DIGICOR
does not properly work.
It may be because the option DIGICOR works for the DIFX data
even for non-VLBA arrays whereas the Daejeon correlator
generates FITS data whose header information is not optimized
for the use of the option DIGICOR.
Since the scaling difference between DIGICOR ON/OFF cases
could be some percent (e.g., 1/0.88 for 2-bit quantization),
the actual difference of the amplitudes between KJCC and DiFX
after applying the option DIGICOR is larger than 8\%.
In order to investigate the difference,
it is required to compare the DIGICOR-ON-outputs of the KVN data
correlated by DiFX and Daejeon correlator with those of the VLBA data
correlated by DiFX.

\section{Observational data and data reduction}

\begin{figure}[!t]
        \centering 
        \includegraphics[width=80mm]{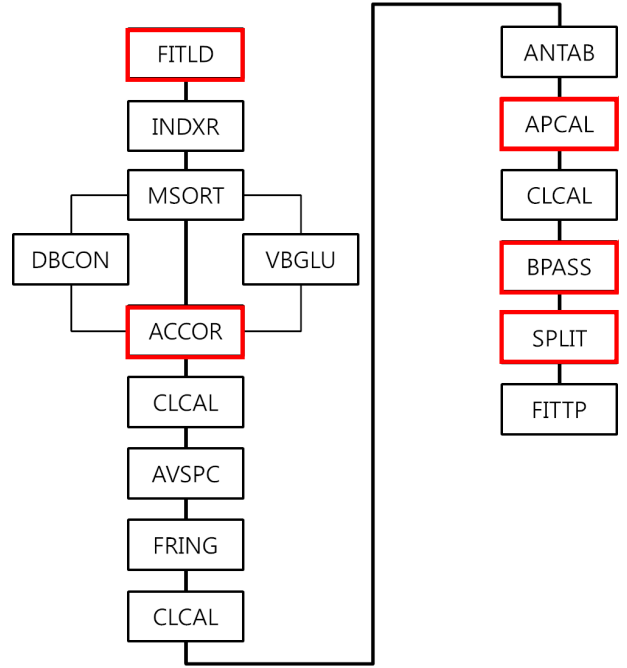}
        \caption{ 
        Schematic diagram of VLBI data reduction with AIPS. Adopted and modified from \cite{lee+15}.
	The AIPS Tasks related with amplitude calibration are outlined in red.
        }
        \label{fig2} 
\end{figure}

\begin{table*}[t!]
\scriptsize
\caption{Log of observational data\label{data}}
\centering
\setlength{\tabcolsep}{0.15cm}
\begin{tabular}{lcccccccc}
\toprule
VLBI & Experiment & Experiment & Frequency Band & Bit rate & Frequcny & Sampling & Bandwidth &  \\
Network & Name       & Date       & (GHz)          &  (bit) & Channels & (Mbit~s$^{-1}$) & (MHz) & Polarization \\
(1) & (2) & (3) & (4) & (5) & (6) & (7) & (8) & (9) \\
\midrule
VLBA& bf104a  & 2011 Jan 17 & 23.50-23.98 &  2 & 8  & 256  & 64 & LCP \\
KVN & r11027b & 2011 Jan 28 & 22.00-22.45 &  2 & 16 & 1024 & 256 & LCP \\
VLBA& bm272   & 2011 Feb 26 & 22.22-22.24 &  2 & 4  & 128  & 32 & Dual \\
VLBA& td074   & 2014 Apr 13 & 22.01-22.11 &  2 & 4  & 512  & 128 & Dual \\
KVN &n14sl01h& 2014 Apr 22 & 21.70-21.76 &  2 & 4  & 1024 & 64 & LCP \\
VLBA&s6096a   & 2014 Apr 26 & 22.09-22.36 &  2 & 8  & 1024 & 256 & Dual \\
\bottomrule
\end{tabular}
\end{table*}

\subsection{Observational data}

We compared KVN data obtained at 22~GHz
with VLBA data at 22~GHz.
We selected the VLBA data obtained close to the KVN data in time
within a month in order to reduce source variability effect.
The observational data at 22 GHz are used since we expect
relatively less uncertainty of the amplitude calibration
at this frequency than at higher frequencies,
due to the atmospheric effect, pointing offset, etc.
We used two KVN experiments (r11027 and n14sl01h) on 2011 January 28
and 2014 April 22
together with four VLBA experiments (BF104A, BM272, TD074, and S6096A)
on 2011 January 17, 2011 February 26, 2014 April 13 and 2014 April 26.
All experiments were conducted at 22 GHz bands.
The KVN and VLBA data for this paper are summarized in Table~\ref{data}.

\subsection{Data reduction}\label{datareduction}

For all of the data sets selected,
we performed the same post-correlation processing using the NRAO Astronomical
Imaging Processing System (AIPS),
following a standard processing procedure as described in Figure~\ref{fig2}.
The correlated output FITS data was uploaded by FITLD task.
In using FITLD, we did apply digital correction (DIGICOR = 1)
for all the data sets whether it works or not.
This is the first task related with the amplitude calibration
and one proper chance to correct for
the quantization loss by the digital samplers. 
As described in Section~\ref{correlator},
for the VLBI data (e.g., the KVN and VLBA data) correlated by 
DiFX, the quantization loss can be corrected by the DIGICOR option,
whereas, for the data correlated by the Daejeon correlator,
the DIGICOR option does not change the visibility amplitude,
hence no correction for the quantization loss.
The loaded data was
indexed with INDXR task,
sorted in an order of time and baseline by MSORT,
combined, if necessary, in time or frequency by DBCON or VBGLU, respectively,
and corrected by ACCOR
for some amplitude errors in the digital sampling.
This is the second task related with the amplitude calibration,
but has nothing to do with the quantization error correction,
although it is a very important task which could generate the amplitude
calibration error if not properly applied.
After applying the correction information from the ACCOR
by CLCAL and fringe-fitting with FRING,
the atmospheric opacity change and
the amplitude errors due to the atmospheric fluctuation,
are corrected by APCAL.
This is the third task related with the amplitude calibration.
System noise temperatures and antenna gains measured at each observatory
are used for converting correlation coefficient to sky brightness and
correcting for the amplitude errors.
Depending on weather conditions and the accuracy of antenna gain measurements,
we may expect the amplitude calibration errors as large as 10-30\%
for any VLBI array.
The amplitude-calibrated data are further corrected for 
effect of bandpass filter on the spectrum shape using BPASS task. 
This is the fourth task related with the amplitude calibration.
In this step, we only calibrated the amplitude with the BPASS task
using the full bandwidth.
For the multiple-source data, we split the data for individual source
by SPLIT with chopping 10\% of frequency channels at the band edge.
This is the fifth step related to the amplitude calibration,
since the excluded channels may be suffering from the amplitude loss
due to the uncorrected effect of the bandpass filter.

\subsection{Amplitude correction}\label{correct}

Any unwanted visibility amplitude offset, which we want to correct,
can be corrected in conducting the APCAL.
The specific APCAL options and parameters for the amplitude correction 
are (a) APARM(1) = $B$, a correction factor (e.g., 1.10), which is called B factor
in AIPS, (b) OPCODE = `', and (c) DOFIT $>$ 1.  
The APCAL with this setup correct the amplitude of the visibility 
by multiplying the correction factor, $B$, with the original amplitude.
It should be noted that since the APCAL corrects the amplitude
of every visibility before the time and frequency averaging,
the resultant scatter of the visibility amplitude may also increase by
a factor of $<B$.

\section{Results}

\begin{table*}[t!]
\scriptsize
\caption{Mean visibility amplitude\label{comp}}
\centering
\setlength{\tabcolsep}{0.15cm}
\begin{tabular}{lcccccccc}
\toprule
       &           &            &          &              &  Length      & P.A.        & Amplitude & Corrected amplitude\\
Source & Telescope & Correlator & Epoch    & Baseline     & (M$\lambda$) & $(^{\circ})$&(Jy)       & (Jy) \\
(1)    &       (2) &        (3) &      (4) &          (5) & (6)          & (7)         & (8)       & (9)  \\
\midrule
3C~454.3& KVN      & DiFX       &2011 Jan 28 & KT-KY & 34 & 20         & $24.7\pm0.7$ & $27.1\pm0.7$\\
        &          &            &            & KT-KU & 26 & 55         & $23.8\pm0.7$ & $26.1\pm0.8$\\
        &          &            &            & KU-KY & 20 & -30        & $25.6\pm0.4$ & $28.2\pm0.5$\\
        & KVN      & Daejeon    &2011 Jan 28 & KT-KY & 34 & 20         & $20.2\pm0.7$ & $27.3\pm1.0$\\
        &          &            &            & KT-KU & 26 & 55         & $19.5\pm0.7$ & $26.4\pm0.9$\\
        &          &            &            & KU-KY & 20 & -30        & $21.1\pm0.6$ & $28.4\pm0.7$\\
        & VLBA     & DiFX       &2011 Feb 26 & FD-PT & 35 & 20         & $28.2\pm0.4$ & - \\
        &          &            &            & KP-PT & 28 & 65         & $26.8\pm0.8$ & - \\
        &          &            &            & LA-PT & 17 & 60         & $28.0\pm0.4$ & - \\\addlinespace
        & KVN      & DiFX       &2014 Apr 22 & KU-KY & 19 & 20         & $9.64\pm0.06$  & $10.6\pm0.06$\\
        &          &            &            & KT-KY & 32 & -20        & $10.4\pm0.08$  & $11.4\pm0.03$\\
        & VLBA     & DiFX       &2014 Apr 13 & LA-PT & 16 & 40         & $10.4\pm0.08$ & - \\
        &          &            &2014 Apr 26 & LA-PT & 18 & 50         & $11.7\pm0.02$ & - \\
        &          &            &            & KP-PT & 31 & 55         & $11.2\pm0.05$ & - \\\addlinespace
NRAO~512& KVN      & DiFX       &2011 Jan 28 & KT-KY & 32 & 40         & $0.98\pm0.05$ & $1.08\pm0.06$\\
        &          &            &            & KU-KY & 22 & 10         & $1.00\pm0.05$ & $1.10\pm0.05$\\
        &          &            &            & KT-KU & 18 & 80         & $1.00\pm0.06$ & $1.10\pm0.07$\\
        & KVN      & Daejeon    &2011 Jan 28 & KT-KY & 32 & 40         & $0.79\pm0.05$ & $1.07\pm0.07$\\
        &          &            &            & KU-KY & 22 & 10         & $0.81\pm0.05$ & $1.09\pm0.06$\\
        &          &            &            & KT-KU & 18 & 80         & $0.80\pm0.05$ & $1.08\pm0.07$\\
        & VLBA     & DiFX       &2011 Jan 17 & KP-PT & 32 & -15        & $1.00\pm0.14$ & - \\
        &          &            &            & FD-LA & 32 & -55        & $1.11\pm0.11$ & - \\
        &          &            &            & FD-KP & 31 & 25          & $1.06\pm0.13$& - \\
        &          &            &            & LA-PT & 18 & -20        & $1.07\pm0.14$ & - \\
OJ~287  & KVN      & DiFX       &2014 Apr 22 & KU-KY & 18 & 5          & $3.74\pm0.01$ & $4.08\pm0.01$\\
        &          &            &            & KT-KY & 33 & 0          & $4.00\pm0.01$ & $4.40\pm0.01$\\
        & VLBA     & DiFX       &2014 Apr 26 & LA-PT & 18 & 50         & $4.43\pm0.01$ & - \\
        &          &            &            & KP-PT & 31 & 55         & $4.22\pm0.02$ & - \\\addlinespace
BL~Lac  & KVN      & DiFX       &2014 Apr 22 & KU-KY & 14 & -80        & $4.98\pm0.01$ & $5.48\pm0.01$\\
        &          &            &            & KU-KY & 17 & -75        & $4.83\pm0.03$ & $5.32\pm0.03$\\
        &          &            &            & KT-KY & 33 & -35        & $4.31\pm0.01$ & $4.54\pm0.02$\\
        & VLBA     & DiFX       &2014 Apr 26 & LA-PT & 14 & 75         & $5.72\pm0.01$ & - \\
        &          &            &            & LA-PT & 17 & 55         & $5.27\pm0.01$ & - \\
        &          &            &            & KP-PT & 31 & 45         & $4.50\pm0.02$ & - \\\addlinespace
3C~279  & KVN      & DiFX       &2014 Apr 22 & KU-KY & 12 & -20        & $24.6\pm0.07$ & $27.1\pm0.08$\\
        &          &            &            & KT-KU & 26 & 55         & $23.3\pm0.03$ & $25.6\pm0.03$\\
        &          &            &            & KT-KY & 28 & 20         & $23.6\pm0.02$ & $26.0\pm0.02$\\
        & VLBA     & DiFX       &2014 Apr 26 & LA-PT & 12 & 45         & $26.0\pm0.07$ & - \\
        &          &            &            & KP-PT & 26 & 55         & $26.3\pm0.2$ & - \\
        &          &            &            & KP-PT & 28 & 45         & $24.7\pm0.1$ & - \\\addlinespace
1633+382& KVN      & DiFX       &2014 Apr 22 & KU-KY & 17 & -70        & $2.70\pm0.01$ & $2.98\pm0.02$\\
        &          &            &            & KT-KY & 33 & -35        & $2.61\pm0.02$ & $2.88\pm0.02$\\
        & VLBA     & DiFX       &2014 Apr 26 & LA-PT & 17 & 30         & $2.99\pm0.01$ & - \\
        &          &            &            & KP-LA & 33 & 80         & $2.61\pm0.02$ & - \\\addlinespace
1510$-$089& KVN    & DiFX       &2014 Apr 22 & KT-KU & 17 & 45         & $3.11\pm0.01$ & $3.42\pm0.01$\\
        &          &            &            & KT-KY & 27 & -20        & $3.20\pm0.02$ & $3.52\pm0.02$\\
        & VLBA     & DiFX       &2014 Apr 26 & LA-PT & 17 & 55         & $3.33\pm0.01$ & - \\
        &          &            &            & KP-OV & 27 & -15        & $3.48\pm0.04$ & - \\
\bottomrule
\end{tabular}
\end{table*}

\begin{table*}[t!]
\scriptsize
\caption{Comparison of the visibility amplitude\label{factor}}
\centering
\setlength{\tabcolsep}{0.15cm}
\begin{tabular}{lcccccc}
\toprule
Source & Epoch & Baseline  & $\frac{S_{\rm VLBA}}{S_{\rm KVN,DiFX}}$ & $\frac{S_{\rm VLBA}}{S_{\rm KVN,Daej}}$  & $\frac{S_{\rm KVN,DiFX}}{S_{\rm KVN,Daej}}$ \\
\midrule
3C~454.3& 2011 Jan 28 & KT-KY    & 1.14   & 1.40  & 1.22 \\
        &             & KU-KY    & 1.13   & 1.37  & 1.22 \\
        &             & KT-KU    & 1.09   & 1.33  & 1.21\\ \addlinespace
        & 2014 Apr 22 & KT-KY    & 1.08   & -     & -   \\
        &             & KU-KY    & 1.09   & -     & -   \\ \addlinespace
NRAO~512& 2011 Jan 28 & KT-KY    & 1.08   & 1.34  & 1.24 \\
        &             & KU-KY    & -      & -     & 1.23 \\
        &             & KT-KU    & 1.07   & 1.34  & 1.25\\ \addlinespace
OJ~287  & 2014 Apr 22 & KT-KY    & 1.06   & -     & -   \\
        &             & KU-KY    & 1.19   & -     & -   \\ \addlinespace
BL~Lac  & 2014 Apr 22 & KT-KY    & 1.04   & -     & -   \\
        &             & KU-KY    & 1.12   & -     & -   \\ \addlinespace
3C~279  & 2014 Apr 22 & KT-KY    & 1.05   & -     & -   \\
        &             & KU-KY    & 1.06   & -     & -   \\
        &             & KT-KU    & 1.13   & -     & -   \\ \addlinespace
1633+382& 2014 Apr 22 & KT-KY    & 1.14   & -     & -   \\
        &             & KU-KY    & 1.11   & -     & -   \\ \addlinespace
1510$-$089&2014 Apr 22& KT-KY    & 1.09   & -     & -   \\
        &             & KU-KY    & 1.08   & -     & -   \\
\hline
Mean    &             &          & 1.10   & 1.35  & 1.23\\
\bottomrule
\end{tabular}
\end{table*}

\begin{table*}[t!]
\scriptsize
\caption{Amplitude correction factors\label{tab}}
\centering
\setlength{\tabcolsep}{0.15cm}
\begin{tabular}{lcccc}
\toprule
Telescope & Correlator & Frequency band  & Correction Factor & Remarks \\
\midrule
KVN       & DiFX       & 22~GHz          & 1.10              & \\
          &            & 43-129~GHz      & {\it 1.10}        & \\
          & Daejeon    & 22~GHz          & 1.35              & before 2015 March\\
          &            & 22~GHz          & 1.30              & after  2015 March\\
          &            & 43-129~GHz      & {\it 1.35}        & before 2015 March\\
          &            & 43-129~GHz      & {\it 1.30}        & after  2015 March\\ \addlinespace
VERA      & DiFX       & 22-43~GHz       & {\it 1.10}        & \\
          & Daejeon    & 22-43~GHz       & {\it 1.35}        & before 2015 March\\
          &            & 22-43~GHz       & {\it 1.30}        & after  2015 March\\
\bottomrule
\end{tabular}
\tabnote{
The correction factors should be applied only to the KVN observations at a recording
rate of 1024~Mbps. The correction factors in italic are the suggested values.
}
\end{table*}

After the amplitude calibration, we selected visibility data
of
3C~454.3, NRAO~512, OJ 287, BL Lac, 3C 279, 1633+382, and 1510$-$089,
obtained with similar baselines
at KVN and VLBA.
Since, at 22~GHz band, the KVN baseline lengths are
in the range of 20-35~M$\lambda$,
we selected the VLBA data on the baselines,
Fort Davis (FD) to Pie Town (PT),
Kitt Peak (KP) to PT, Los Alamos (LA) to PT, FD to LA, FD to KP,
or KP to Owens Valley (OV), 
whose baseline lengths are in the range of 17-35~M$\lambda$.
The visibility data were first averaged in time at an interval of 30~s.
The amplitudes of the visibility data for each baseline
were averaged for all IF bands.
The estimated mean visibility amplitudes for all baselines and sources
are summarized in Table~\ref{comp}.
The estimated mean amplitudes are compared between the KVN and VLBA baselines
with similar baseline lengths and also compared between the KVN-DiFX
and the KVN-Dajeon data.
By taking into the comparison results, we determined the amplitude
correction factors for the KVN observations correlated by DiFX
and Daejeon correlators.

In the KVN baselines (or resolutions),
the source compactness (defined as the ratio of
the correlated flux density on the longest baseline to that on the shortest baseline)
is close to unity, implying that the source is very compact or unresolved.
Since some of the VLBA baselines selected are not aligned with
the KVN baselines in the position angle (P.A.) of the baseline,
we may expect that a source structure resolved in the KVN baselines
may results in different correlated flux density depending on 
the position angle of the baselines. 
However, we expect that the source structure effects
are very small due to the compact structure of the sources.
We found that, on average, the visibility amplitudes of VLBA data
are higher than those of KVN-DiFX data by 10\%
and much higher than those of KVN-Daejeon data by 35\%, as shown
in Table~\ref{factor}.
This is because the visibility amplitudes of the KVN-DiFX data
are higher than those of the KVN-Daejeon by 23\%.

By the amplitude correction factors determined,
we corrected the amplitude of the KVN observations according to
Section~\ref{correct}.
As shown in Table~\ref{comp} (Column 8),
the mean visibility amplitudes for the KVN data are corrected to be consistent
to those of the VLBA data.
The scatter of the visibility amplitude get larger by the applied factors
as expected.

\section{Discussion}

The difference of the calibrated visibility amplitudes
between KVN-DiFX, KVN-Daejeon, and VLBA-DiFX data
are attributed to several aspects.
We will consider here the factors which are
important in causing the amplitude difference,
and determine the correction factors for the KVN amplitude calibration.

\subsection{Quantization loss}

The quantization loss in digitizing analog signals
has been theoretically analyzed~\citep[e.g.,][]{tho+01},
and the digital loss in re-quantizing the digitized signals
has been numerically investigated by \cite{igu+05}.
The quantization loss is known to be 0.88 for the four-level (2-bit)
quantization
and the accumulated digital loss was shown to be 0.81 ($=0.88\times0.92$)
in four-level re-quantization of the four-level quantized signals.
Thus the additional digital loss due only to the four-level re-quantization
is 0.92 in the case of the input signals of four-level quantization.

KVN observations experience a four-level quantization
in the digital samplers and an additional four-level re-quantization in the DFB.
As described in Section~\ref{digital},
the digital sampling and filtering are based on the four-level (2-bit)
quantization, resulting in an accumulated digital loss factor of 0.81. 
In a standard calibration procedure, for example as described
in Section~\ref{datareduction},
a quantization loss factor of 0.88 is corrected by switching
on the option DIGICOR of the FITLD task.
Thus, the quantization loss by the digital samplers
can be corrected with the standard calibration procedure,
whereas the re-quantization loss by the DFB is not, in general procedure. 
Consequently, the remaining amplitude loss factor
for the properly calibrated visibility of the DiFX correlator
is about 0.92, and the correction factor is 1.09,
which is similar to the amplitude correction factor
$\frac{S_{\rm VLBA}}{S_{\rm KVN,DiFX}}=1.10$.
The difference between the two correction factors, 1.09 and 1.10,
may come from the difference of the amplitude calibration uncertainties
of KVN and VLBA (see below).

\subsection{Characteristics of Daejeon correlator}

As reported by \cite{lee+15}, the calibrated visibility of
the Daejeon correlator suffers from two amplitude loss factors:
(a) 0.95 (5\%)  due to the scheme of fringe tracking
and (b) 0.97 (3\%) due to the double-layer pattern problem.
The fringe tracking scheme is the characteristics
of the VLBI Correlation System (VCS) of the hardware correlator,
which should be a constant loss factor
as long as the hardware correlator uses the same fringe tracking scheme
in the VCS. 
The double-layer pattern problem turns out to happen 
in serializing the data and mapping the memory
in the VCS,
and disappears in correlations after 2015 March
(S.~J. Oh, priv. communication). 

In addition to these loss factors, 
the calibrated visibility of the Daejeon correlator
may experience additional amplitude loss
after the proper amplitude calibration 
since the quantization and re-quantization loss from the KVN observations
is not properly corrected using a standard VLBI
data reduction procedure,
as described in Section~\ref{correlator}.
This characteristics results in an amplitude loss factor of 0.81. 
Therefore, the total amplitude loss factor
for the properly calibrated visibility of the Daejeon correlator
is about 0.75 ($=0.81\times0.95\times0.97$), and the correction factor is 1.34,
which is similar to the amplitude correction factor
$\frac{S_{\rm VLBA}}{S_{\rm KVN,Daej}}=1.35$.
The difference between the two correction factors, 1.34 and 1.35,
may come from the difference of the amplitude calibration uncertainties
of KVN and VLBA (see below).
Moreover, the amplitude ratio
$\frac{S_{\rm KVN, DiFX}}{S_{\rm KVN,Daej}}=1.23$
can be explained since the total amplitude
loss factor of the Daejeon correlator compared to the DiFX
is about 0.81 ($=0.88\times0.92$)
and hence the correction factor is 1.23.

\subsection{Calibration uncertainty and source variability}

In addition to the two factors, the amplitude difference may also be
attributed to the amplitude calibration uncertainties of KVN and VLBA
and to the time-variability of the source brightness and structure.
The difference of the amplitude calibration uncertainties
may depend on the observing frequency band and the weather condition.
And the source variability effect may be reduced
by using the contemporaneous data from KVN and VLBA.
Therefore, these two factors contribute very less to the difference
of the visibility amplitude between KVN and VLBA data than
the factors described above.
Caution should be taken here,
as the weather conditions are totally different in KVN and VLBA,
and thus any systematics which could be caused by the weather
may not be negligible.

\subsection{Amplitude correction factors}

Since the differences of the calibrated visibility amplitudes
between KVN and VLBA data are attributed mainly to 
the digital characteristics of the KVN VLBI system
and the hardware and software characteristics of Daejeon correlator,
we may expect that the nature of the difference should remain constant
in time and frequency bands.
Therefore, one may apply the amplitude correction factors 
to the future KVN observations at 22~GHz, 43~GHz, 86~GHz and 129~GHz bands.
The time and frequency dependence of the amplitude correction factors
will be extensively investigated by using the KVN data of a KVN key science program,
iMOGABA (Interferometric Monitoring of Gamma-ray Bright AGNs)~\citep[][Lee et al. in prep.]{waj+15,alg+15},
and the results will be reported elsewhere.

Our investigations are based on the KVN observations.
However, the results may be applied to the VERA observations,
since the backend system (in particular, digital sampler and DFB), of the VERA 
is similar to that of the KVN.
Thus, we suggest that the amplitude correction factors
determined in this paper can be applied to the VERA or KVN-VERA observations.
Again, the amplitude correction factors for the KVN-VERA data
will be confirmed and reported elsewhere (Oh. et al. in prep.).

Finally, we suggest to use the amplitude correction factors of
1.30 ($=\frac{1}{0.81\times0.95})$ for
the VLBI data correlated by the Daejeon correlator since 2015 March,
because the double-layer pattern problem is announced to be resolved
for the recent correlations.  
All the determined and suggested correction factors are summarized 
in Table~\ref{tab}.
The correction factors should be applied only to the KVN observations
at a recording rate of 1024~Mbps.

\section{Conclusion}

The basic assumption that the VLBA amplitude calibration is robust
enables to evaluate the amplitude calibration of KVN observations
based on the rigorous comparison between the KVN and VLBA data
obtained contemporaneously for extragalactic compact radio sources.
We found that the VLBA data have higher amplitude
than the KVN data by factors of 1.10 and 1.35 for the cases
of DiFX and Daejeon correlators.
Among several aspects contributing to the amplitude difference,
we found that the quantization losses by the digital samplers
and the DFB, and the characteristics of the Daejeon correlator
are the main contributors.
We expect that the nature of the amplitude losses is constant in time and frequency band.
Therefore, we should apply the amplitude correction factors of 1.10 and
1.35 to the future KVN observations correlated by DiFX and Daejeon correlators,
respectively.
It is also suggested to apply these correction factors to observations
using the KVN and VERA combined array. 


\acknowledgments

We would like to thank the anonymous referee for important comments
and suggestions which have enormously improved the manuscript.
We are grateful to all staff members in KVN
who helped to operate the array and to correlate the data.
The KVN is a facility operated by
the Korea Astronomy and Space Science Institute.
The KVN operations are supported
by KREONET (Korea Research Environment Open NETwork)
which is managed and operated
by KISTI (Korea Institute of Science and Technology Information).
The VLBA is an instrument of the National Radio Astronomy Observatory,
which is a facility of the National Science Foundation operated under
cooperative agreement by Associated Universities, Inc.

\end{document}